\documentclass[11pt]{revtex4}    % updated 7 May 2010%.
\usepackage{amssymb,epsf}
\usepackage{latexsym}
\begin{document}
\title{Thermodynamics of apparent horizon and modified Friedman equations}
\author{Ahmad Sheykhi\footnote{sheykhi@mail.uk.ac.ir}}
\address{Department of Physics, Shahid Bahonar University, P.O. Box 76175, Kerman, Iran\\
         Research Institute for Astronomy and Astrophysics of Maragha (RIAAM), Maragha,
        Iran}

 \begin{abstract}
\vspace*{1.5cm} \centerline{\bf Abstract} \vspace*{1cm} Starting
from the first law of thermodynamics, $dE=T_hdS_h+WdV$, at
apparent horizon of a FRW universe, and assuming that the
associated entropy with apparent horizon has a quantum corrected
relation, $S=\frac{A}{4G}-\alpha \ln \frac{A}{4G}+\beta
\frac{4G}{A}$, we derive modified Friedmann equations describing
the dynamics of the universe with any spatial curvature.  We also
examine the time evolution of the total entropy including the
quantum corrected entropy associated with the apparent horizon
together with the matter field entropy inside the apparent
horizon. Our study shows that, with the local equilibrium
assumption, the generalized second law of thermodynamics is
fulfilled in a region enclosed by the apparent horizon.

\end{abstract}
 \maketitle

 \newpage
\section{Introduction\label{Intro}}
The pioneer study on the deep connection between gravity and
thermodynamics was done by Jacobson \cite{Jac} who showed that the
gravitational Einstein equation can be derived from the relation
between the horizon area and entropy, together with the Clausius
relation $\delta Q=T\delta S$. Further studies on the connection
between gravity and thermodynamics has been investigated in
various gravity theories \cite{Elin,Cai1,Pad}. In the cosmological
context, attempts to disclose the connection between Einstein
gravity and thermodynamics were carried out in
\cite{Cai2,Cai3,CaiKim,Fro,verlinde}. It was shown that the
differential form of the Friedmann equation in the
Friedmann-Robertson-Walker (FRW) universe can be written in the
form of the first law of thermodynamics on the apparent horizon.
The profound connection provides a thermodynamical interpretation
of gravity which makes it interesting to explore the cosmological
properties through thermodynamics. Investigations on the deep
connection between gravity and thermodynamics have recently been
extended to braneworld scenarios \cite{Cai4,Shey1,Shey2}.

It is interesting to note that Friedmann equations, in Einstein's
gravity, can be derived by applying the Clausius relation to the
apparent horizon of FRW universe, in which entropy is assumed to
be proportional to its horizon area, $S={A}/{4G}$ \cite{CaiKim}.
However, this definition for entropy can be modified from the
inclusion of quantum effects, motivated from the loop quantum
gravity (LQG). The quantum corrections provided to the
entropy-area relationship leads to the curvature correction in the
Einstein-Hilbert action and vice versa \cite{Zhu}. The corrected
entropy takes the form \cite{Zhang}
\begin{equation}\label{S}
S_h=\frac{A}{4G}-\alpha \ln \frac{A}{4G}+\beta \frac{4G}{A},
\end{equation}
where $\alpha$ and $\beta$ are positive dimensionless constants of
order unity. The exact values of these constants are not yet
determined and still an open issue in loop quantum cosmology.
These corrections arise in the black hole entropy in LQG due to
thermal equilibrium fluctuations and quantum fluctuations
\cite{Rovelli}. It is important to note that in the literature
different kind of modification of entropy expression have studied
in classical level for various modified gravity theories
\cite{Nojiri1,Nojiri11}. The log correction to the area-entropy
relation appears to have an almost universal status, having been
derived from multiple different approaches to the calculation of
entropy from counting microscopic states in different quantum
gravity models. Let us stress here that although in the literature
there is doubt about the second correction term in
entropy-corrected relation, however, it is  widely believed
\cite{Zhang} that the next quantum correction term to black hole
entropy have the form ${4G}/{A}$, which leads to the reasonable
correction terms to Newton's law of gravitation \cite{shey0} and
will also lead to the corrected modified Friedmann equation as we
will show in this paper.

Besides, if thermodynamical interpretation of gravity near
apparent horizon is generic feature, one needs to verify whether
the results may hold not only for more general spacetimes but also
for the other principles of thermodynamics, especially for the
generalized second law of thermodynamics. The generalized second
law of thermodynamics is a universal principle governing the
universe. The generalized second law of thermodynamics in the
accelerating universe enveloped by the apparent horizon has been
studied extensively in \cite{wang1,wang2,Shey3}. For other gravity
theories, the generalized second law has also been considered in
\cite{akbar}.

The aim of this paper is twofold. The first is to derive modified
Friedmann equations by applying  the first law of thermodynamics,
$dE=T_hdS_h+WdV$, at apparent horizon of a FRW universe and
assuming the apparent horizon has an entropy expression like (1).
The other is to see whether the quantum corrected entropy-area
relation together with the matter field entropy inside the
apparent horizon will satisfy the generalized second law of
thermodynamics.

\section{Modified Friedmann Equation from the First law of thermodynamics\label{FIRST}}
We consider a homogenous and isotropic FRW universe which is
described by the line element
\begin{equation}
ds^2={h}_{\mu \nu}dx^{\mu}
dx^{\nu}+\tilde{r}^2(d\theta^2+\sin^2\theta d\phi^2),
\end{equation}
where $\tilde{r}=a(t)r$, $x^0=t, x^1=r$, the two dimensional
metric $h_{\mu \nu}$=diag $(-1, a^2/(1-kr^2))$. Here $k$ denotes
the curvature of space with $k = 0, 1, -1$ corresponding to open,
flat, and closed universes, respectively. The dynamical apparent
horizon, a marginally trapped surface with vanishing expansion, is
determined by the relation $h^{\mu \nu}\partial_{\mu}\tilde
{r}\partial_{\nu}\tilde {r}=0$. Straightforward calculation gives
the apparent horizon radius for the FRW universe
\begin{equation}
\label{radius}
 \tilde{r}_A=\frac{1}{\sqrt{H^2+k/a^2}}.
\end{equation}
The associated temperature with the apparent horizon can be
defined as $T = \kappa/2\pi$, where $\kappa$ is the surface
gravity $
 \kappa =\frac{1}{\sqrt{-h}}\partial_{\mu}\left(\sqrt{-h}h^{\mu \nu}\partial_{\mu\nu}\tilde
 {r}\right).
$ Then one can easily show that the surface gravity at the
apparent horizon of FRW universe can be written as
\begin{equation}\label{surgrav}
 \kappa=-\frac{1}{\tilde
r_A}\left(1-\frac{\dot {\tilde r}_A}{2H\tilde r_A}\right).
\end{equation}
When $\dot {\tilde r}_A\leq 2H\tilde r_A$, the surface gravity
$\kappa\leq 0$, which leads the temperature $T\leq 0$ if one
defines the temperature of the apparent horizon as $T=\kappa/2\pi$
. Physically it is not easy to accept the negative temperature,
the temperature on the apparent horizon should be defined as
$T=|\kappa|/2\pi$. Recently, the connection between temperature on
the apparent horizon and the Hawking radiation has been considered
in \cite{cao}, which gives more solid physical implication of the
temperature associated with the apparent horizon.

Suppose the matter source in the FRW universe is a perfect fluid
with stress-energy tensor
\begin{equation}\label{T}
T_{\mu\nu}=(\rho+p)u_{\mu}u_{\nu}+pg_{\mu\nu},
\end{equation}
where $\rho$ and $p$ are the energy density and pressure,
respectively. The energy conservation law then leads to
\begin{equation}\label{Cont}
\dot{\rho}+3H(\rho+p)=0,
\end{equation}
where $H=\dot{a}/a$ is the Hubble parameter. Following
\cite{Hay2}, we define the work density as
\begin{equation}\label{Work}
W=-\frac{1}{2} T^{\mu\nu}h_{\mu\nu}.
\end{equation}
In our case it becomes
\begin{equation}\label{Work2}
W=\frac{1}{2}(\rho-p).
\end{equation}
The work density term is regarded as the work done by the change
of the apparent horizon. Assuming the first law of thermodynamics
on the apparent horizon is satisfied and has the form
\begin{equation}\label{FL}
dE = T_h dS_h + WdV,
\end{equation}
where $S_{h}$ is the quantum corrected entropy associated with the
apparent horizon which has the form (1). One can see that in Eq.
(\ref{FL}), the work density is  replaced with the negative
pressure if we compare with the standard first law of
thermodynamics, $dE = TdS-pdV$. For a pure de Sitter space,
$\rho=-p$, then the work term reduces to the standard $-pdV$ and
we obtain exactly the standard first law of thermodynamics. \\ We
also assume $E=\rho V$ is the total energy content of the universe
inside a $3$-sphere of radius $\tilde{r}_{A}$, where
$V=\frac{4\pi}{3}\tilde{r}_{A}^{3}$ is the volume enveloped by
3-dimensional sphere with the area of apparent horizon
$A=4\pi\tilde{r}_{A}^{2}$. Taking differential form of the
relation $E=\rho \frac{4\pi}{3}\tilde{r}_{A}^{3}$ for the total
matter and energy inside the apparent horizon, we get
\begin{equation}
\label{dE1}
 dE=4\pi\tilde
 {r}_{A}^{2}\rho d\tilde {r}_{A}+\frac{4\pi}{3}\tilde{r}_{A}^{3}\dot{\rho} dt.
\end{equation}
Using the continuity equation (\ref{Cont}), we obtain
\begin{equation}
\label{dE2}
 dE=4\pi\tilde
 {r}_{A}^{2}\rho d\tilde {r}_{A}-4\pi H \tilde{r}_{A}^{3}(\rho+p) dt.
\end{equation}
Taking differential form of the corrected entropy (\ref{S}), we
have
\begin{equation} \label{dS}
dS_h= \frac{2\pi \tilde {r}_{A}}{G}\left[1-\frac{\alpha G}{\pi
{\tilde {r}_{A}}^2}-\frac{\beta G^2}{\pi^2 {\tilde
{r}_{A}}^4}\right] d\tilde {r}_{A}.
\end{equation}
Inserting Eqs. (\ref{Work2}), (\ref{dE2}) and (\ref{dS}) in the
first law (\ref{FL}) and using the relation between temperature
and surface gravity, we can get the differential form of the
modified Friedmann equation
\begin{equation} \label{Fried1}
\frac{1}{4\pi G}\frac{d\tilde {r}_{A}}{\tilde
{r}_{A}^3}\left[1-\frac{\alpha G}{\pi {\tilde
{r}_{A}}^2}-\frac{\beta G^2}{\pi^2 {\tilde {r}_{A}}^4}\right] = H
(\rho+p) dt.
\end{equation}
Using the continuty equation (\ref{Cont}), we can rewrite it as
\begin{equation} \label{Fried2}
\frac{-2d\tilde {r}_{A}}{\tilde {r}_{A}^3}\left[1-\frac{\alpha
G}{\pi {\tilde {r}_{A}}^2}-\frac{\beta G^2}{\pi^2 {\tilde
{r}_{A}}^4}\right] = \frac{8\pi G}{3}d\rho.
\end{equation}
Integrating (\ref{Fried2}) yields
\begin{equation} \label{Fried3}
\frac{1}{{\tilde {r}_{A}}^2}-\frac{\alpha G}{2\pi {\tilde
{r}_{A}}^4}-\frac{\beta G^2}{3\pi^2 {\tilde {r}_{A}}^6}=
\frac{8\pi G}{3}\rho,
\end{equation}
where an integration constant, which is just the cosmological
constant, has been absorbed into the energy density $\rho$.
Substituting $\tilde {r}_{A}$ from Eq.(\ref{radius}) we obtain
entropy-corrected Friedmann equation
\begin{equation} \label{Fried4}
H^2+\frac{k}{a^2}-\frac{\alpha G}{2\pi
}\left(H^2+\frac{k}{a^2}\right)^2-\frac{\beta
G^2}{3\pi^2}\left(H^2+\frac{k}{a^2}\right)^3 = \frac{8\pi
G}{3}\rho.
\end{equation}
In this way we derive the entropy-corrected Friedmann equation by
starting from the first law of thermodynamics, $dE=T_hdS_h+WdV$,
at apparent horizon of a FRW universe, and assuming that the
associated entropy with apparent horizon has a quantum corrected
relation (1). In the absence of the correction terms
$(\alpha=0=\beta)$, one recovers the well-known Friedmann equation
in standard cosmology. Since the last two terms in Eq.
(\ref{Fried4}) can be comparable to the first term only when $a$
is very small, the corrections make sense only at early stage of
the universe where $a\rightarrow 0$. When the universe becomes
large, the entropy-corrected Friedmann equation reduces to the
standard Friedmann equation.

It is important to note that in the literature many different
modifications of entropy and therefore of Friedmann equations are
studied in classical modified gravity theories. For example, in
\cite{Nojiri2} the modified gravity with $\ln R$ or $R^{-n}$ terms
which grow at small curvature was discussed. It was shown
\cite{Nojiri2} that such a model may eliminate the need for dark
energy and may provide the current cosmic acceleration. It was
also demonstrated that $R^2$ terms are important not only for
early time inflation but also to avoid the instabilities and the
linear growth of the gravitational force. Thus, modified gravity
with $R^2$ term seems to be viable classical theory. It was also
argued in \cite{Nojiri11,Nojiri3} that the modified gravity where
some arbitrary function of Gauss-Bonnet term is added to Einstein
action can explain the dark energy dominated universe. It was
shown that such theory may pass solar system tests and can
describe the most interesting features of late-time cosmology such
as the transition from deceleration to acceleration, crossing the
phantom divide, current acceleration with effective (cosmological
constant, quintessence or phantom) equation of state of the
universe. In \cite{Nojiri4} the modification of the Friedmann
equations which may be caused by $f(R)$ gravity, string-inspired
scalar-Gauss-Bonnet, modified Gauss-Bonnet theories, and ideal
fluid with the inhomogeneous equation of state. It was
demonstrated \cite{Nojiri4} that the history of the expansion of
the universe can be reconstructed through such a universal
formulation. Further investigations on the cosmological
implications of the modified theory of gravity have been carried
out in \cite{Nojiri5}.

It is also worth mentioning that Eq. (\ref{Fried4}) is in complete
agreement with the result of \cite{Cai5}. However, our derivation
is quite different from \cite{Cai5}. Let us stress the difference
between here and \cite{Cai5}. First of all, the authors of
\cite{Cai5} have derived modified Friedmann equations by applying
the first law of thermodynamics, $TdS =-dE$, to the apparent
horizon of a FRW universe with the assumption that the apparent
horizon has temperature $T= 1/2\pi\tilde{r}_{A}$ and
corrected-entropy like (1). It is worthy to note  that the
notation $dE$ in \cite{Cai5} is quite different from the same we
used in the present work. In \cite{Cai5}, $-dE$ is actually just
the heat flux $\delta Q$ in \cite{Jac} crossing the apparent
horizon within an infinitesimal internal of time $dt$. But, here
$dE$ is change in the the matter energy inside the apparent
horizon. Besides, in \cite{Cai5} the apparent horizon radius
$\tilde{r}_{A}$ has been assumed to be fixed. Thus, the
temperature of apparent horizon can be approximated to $T=
1/2\pi\tilde{r}_{A}$ and there is no the term of volume change in
it. But, here, we have used the matter energy $E$ inside the
apparent horizon and the apparent horizon radius changes with
time. This is the reason why we have included the term $WdV$ in
the first law (\ref{FL}). Indeed, the term $4\pi\tilde
 {r}_{A}^{2}\rho d\tilde {r}_{A}$ in Eq.  (\ref{dE2}) contributes to the
work term, while this term is absent in $dE$ of \cite{Cai5}. This
is consistent with the fact that in thermodynamics the work is
done when the volume of the system is changed. We have assumed
that $d\tilde{r}_{A}$ is the infinitesimal change in the radius of
the apparent horizon in a small time interval $dt$ which causes a
small change $dV$ of volume inside the apparent horizon. Since the
matter energy $E$ is directly related to the radius of the
apparent horizon, therefore, the change of apparent horizon radius
will change the energy $dE$ inside the apparent horizon.

%%%%%%%%%%%%%%%%%%%%%%%%%%%%%%%%%%%%%%%%%%%%%%%%%%%%%%%%%%%%%%%%%%%%%%%%
\section{Generalized Second law of thermodynamics\label{GSL}}
In this section we turn to investigate the validity of the
generalized second law of thermodynamics in  a region enclosed by
the apparent horizon. Differentiating Eq. (\ref{Fried3}) with
respect to the cosmic time and using Eq. (\ref{Cont}) we get
\begin{equation} \label{GSL1}
\frac{-2\dot{\tilde {r}_{A}}}{\tilde
{r}_{A}^3}\left[1-\frac{\alpha G}{\pi {\tilde
{r}_{A}}^2}-\frac{\beta G^2}{\pi^2 {\tilde {r}_{A}}^4}\right]
=-8\pi G H (\rho+p).
\end{equation}
Solving for $\dot{\tilde {r}_{A}}$ we find
\begin{equation} \label{dotr1}
\dot{\tilde {r}_{A}}=4\pi G H \tilde
{r}_{A}^3(\rho+p)\left[1-\frac{\alpha G}{\pi {\tilde
{r}_{A}}^2}-\frac{\beta G^2}{\pi^2 {\tilde
{r}_{A}}^4}\right]^{-1}.
\end{equation}
One can see from the above equation that $\dot{\tilde{r}}_{A}>0$
provided the dominant energy condition, $\rho+p>0$, holds. Let us
now turn to find out $T_{h} \dot{S_{h}}$:
\begin{equation}\label{TSh1}
T_{h} \dot{S_{h}} =\frac{1}{2\pi \tilde r_A}\left(1-\frac{\dot
{\tilde r}_A}{2H\tilde r_A}\right)\frac{d}{dt}
\left(\frac{A}{4G}-\alpha \ln \frac{A}{4G}+\beta
\frac{4G}{A}\right).
\end{equation}
After some simplification and using Eq. (\ref{dotr1}) we obtain
\begin{equation}\label{TSh2}
T_{h} \dot{S_{h}} =4\pi H {\tilde{r}_{A}^3}
(\rho+p)\left(1-\frac{\dot {\tilde r}_A}{2H\tilde r_A}\right).
\end{equation}
In the accelerating universe the dominant energy condition may
violate, $\rho+p<0$, indicating that the second law of
thermodynamics ,$\dot{S_{h}}\geq0$, does not hold. However, as we
will see below the generalized second law of thermodynamics,
$\dot{S_{h}}+\dot{S_{m}}\geq0$, is still fulfilled throughout the
history of the universe. From the Gibbs equation we have
\cite{Pavon2}
\begin{equation}\label{Gib2}
T_m dS_{m}=d(\rho V)+pdV=V d\rho+(\rho+p)dV,
\end{equation}
where $T_{m}$ and $S_{m}$ are, respectively, the temperature and
the entropy of the matter fields inside the apparent horizon. We
limit ourselves to the assumption that the thermal system bounded
by the apparent horizon remains in equilibrium so that the
temperature of the system must be uniform and the same as the
temperature of its boundary. This requires that the temperature
$T_m$ of the energy inside the apparent horizon should be in
equilibrium with the temperature $T_h$ associated with the
apparent horizon, so we have $T_m = T_h$\cite{Pavon2}. This
expression holds in the local equilibrium hypothesis. If the
temperature of the fluid differs much from that of the horizon,
there will be spontaneous heat flow between the horizon and the
fluid and the local equilibrium hypothesis will no longer hold.
Therefore from the Gibbs equation (\ref{Gib2}) we can obtain
\begin{equation}\label{TSm2}
T_{h} \dot{S_{m}} =4\pi {\tilde{r}_{A}^2}\dot {\tilde
r}_A(\rho+p)-4\pi {\tilde{r}_{A}^3}H(\rho+p).
\end{equation}
To check the generalized second law of thermodynamics, we have to
examine the evolution of the total entropy $S_h + S_m$. Adding
equations (\ref{TSh2}) and (\ref{TSm2}),  we get
\begin{equation}\label{GSL2}
T_{h}( \dot{S_{h}}+\dot{S_{m}})=2\pi{\tilde r_A}^{2}(\rho+p)\dot
{\tilde r}_A=\frac{A}{2}(\rho+p) \dot {\tilde r}_A.
\end{equation}
where $A$ is the apparent horizon area. Substituting $\dot {\tilde
r}_A$ from Eq. (\ref{dotr1}) into (\ref{GSL2}) we find
\begin{equation}\label{GSL3}
T_{h}( \dot{S_{h}}+\dot{S_{m}})=2\pi G A H {\tilde
r_A}^{3}(\rho+p)^2 \left[1-\frac{\alpha G}{\pi {\tilde
{r}_{A}}^2}-\frac{\beta G^2}{\pi^2 {\tilde
{r}_{A}}^4}\right]^{-1}.
\end{equation}
It is important to note that the expression in the bracket of Eq.
(24) is positive at the present time, i.e.,
\begin{equation}
\left[1-\frac{\alpha G}{\pi {\tilde {r}_{A}}^2}-\frac{\beta
G^2}{\pi^2 {\tilde {r}_{A}}^4}\right] >0.
\end{equation}
This is due to the fact that at the present time ${\tilde
{r}_{A}}\gg 1 $ while $\alpha\sim O(1)$, $\beta\sim O(1)$ and
$G\sim 10^{-11}$, thus $ \frac{\alpha G}{\pi {\tilde
{r}_{A}}^2}\ll 1$ and $\frac{\beta G^2}{\pi^2 {\tilde
{r}_{A}}^4}\ll1$. At the early time where $\tilde
{r}_{A}\rightarrow 0$ the generalized second law of thermodynamics
may be violated but in that case the local equilibrium hypothesis
is failed too. Besides, from the physical point of view, the
effect of the correction terms on the entropy should be less than
uncorrected term. Thus, the second and third terms on the right
hand side of Eqs. (1) and (12) should be much smaller than the
first term, otherwise these terms cannot be regarded as the
correction terms. For all above reasons we can expand the right
hand side of Eq. (\ref{GSL3}), up to the linear order of $\alpha$
and $\beta$,
\begin{equation}\label{GSL4}
T_{h}( \dot{S_{h}}+\dot{S_{m}})=2\pi G A H {\tilde
r_A}^{3}(\rho+p)^2 \left[1+\frac{\alpha G}{\pi {\tilde
{r}_{A}}^2}+\frac{\beta G^2}{\pi^2 {\tilde {r}_{A}}^4}\right].
\end{equation}
The right hand side of the above equation cannot be negative
throughout the history of the universe, which means that $
\dot{S_{h}}+\dot{S_{m}}\geq0$ always holds. This indicates that
for a universe with any spacial curvature the generalized second
law of thermodynamics is fulfilled in a region enclosed by the
apparent horizon.
\section{Conclusions\label{Con}}
In summary, applying the first law of  thermodynamics,
$dE=T_hdS_h+WdV$, to apparent horizon of a FRW universe with any
spatial curvature and assuming that the apparent horizon has
temperature $T=\frac{1}{2\pi\tilde r_A}\left(1-\frac{\dot {\tilde
r}_A}{2H\tilde r_A}\right)$, and a quantum corrected entropy-area
relation, $S_h=\frac{A}{4G}-\alpha \ln \frac{A}{4G}+\beta
\frac{4G}{A}$, we are able to derive modified Friedmann equations
governing the dynamical evolution of the universe. We have also
investigated the validity of the generalized second law of
thermodynamics for the FRW universe with any spatial curvature. We
have shown that, when thermal system bounded by the apparent
horizon remains in equilibrium with its boundary such that $T_m =
T_h$, the generalized second law of thermodynamics is fulfilled in
a region enclosed by the apparent horizon. The validity of the
generalized second law of thermodynamics for quantum corrected
entropy area relation further supports the thermodynamical
interpretation of gravity and provides more confidence on the
profound connection between gravity and thermodynamics.

It is worth noting that although we derived modified Friedmann
equations corresponding to the corrected entropy-area relation (1)
by applying the first law of thermodynamics to apparent horizon,
it would be of great interest to see whether one is able to get
modified Einstein field equation by following Jacobson argument
\cite{Jac}. This study is of great interest and further shows that
given a thermodynamical relation between entropy and geometry, one
is able to derive corresponding modified Einstein field equation,
showing an interesting connection between them.

Finally, we would like to mention that the higher order terms of
$(H^2+k/a^2)$ in the modified Friedmann equations (16) only
becomes important at early time of the universe. They may
influence the number of e-folds of inflation, or they may give
corrected upper bound on the number of e-folds following the
holographic principle. These should be examined carefully.  Eq.
(16) does not look to influence the late time cosmology. The
detail of this study will be addressed elsewhere.

%%%%%%%%%%%%%%%%%%%%%%%%%%%%%%%%%%%%%%%%%%%%%%%%%%%%%%%%%%%%%%%%%%%%%%%
\acknowledgments{I thank the anonymous referees for constructive
comments. I am also grateful to Prof. B. Wang for helpful
discussions. This work has been supported by Research Institute
for Astronomy and Astrophysics of Maragha, Iran.}

\end{document}